\documentclass[aps,nofootinbib,twocolumn,groupedaddress]{revtex4}
\usepackage[pdftex]{graphicx}
\usepackage{amsmath,amssymb,bm}
\usepackage[mathscr]{eucal}
\usepackage[pdftex]{hyperref}

\newif\ifarxiv
\arxivfalse

\begin		{document}
\def\Nfour	{\mathcal N\,{=}\,4}

\def\Nc		{N_{\rm c}}

\def\t		{\tau}	

\def\half	{{\textstyle \frac 12}}
\def\coeff	#1#2{{\textstyle \frac{#1}{#2}}}

\def\Arxiv      #1 [#2]{\href{http://arxiv.org/abs/#1}{{\tt arXiv:#1 [#2]}}\,}

\title
    {
    Boost invariant flow, black hole formation,
    and far-from-equilibrium dynamics in\\
    \boldmath $\mathcal N = 4$~supersymmetric Yang-Mills theory
    }

\author{Paul~M.~Chesler\footnotemark[1]}
\author{Laurence~G.~Yaffe\footnotemark[2]}

\affiliation
    {Department of Physics, University of Washington, Seattle, WA 98195, USA}

\date{\today}

\begin{abstract}
Using gauge/gravity duality, we study the creation and evolution
of boost invariant anisotropic,
strongly coupled $\Nfour$ supersymmetric Yang-Mills plasma.
In the dual gravitational description, this corresponds to horizon formation
in a geometry driven to be anisotropic by a time-dependent change
in boundary conditions.
\end{abstract}


\maketitle
\iftrue

\footnotetext[1]{Email: \tt pchesler@u.washington.edu}
\footnotetext[2]{Email: \tt yaffe@phys.washington.edu}

\fi

\section{{Introduction}}

The study of non-equilibrium  phenomena in QCD and other
non-Abelian quantum field theories is a topic of much interest,
with applications to heavy-ion collisions, early universe cosmology,
and other areas.
Much has been learned about near-equilibrium dynamics at weak coupling 
where a quasiparticle picture is valid
\cite{Arnold:2002zm, Arnold:2003zc, Arnold:2004ti,Arnold:2000dr,Jeon:1995zm},
and there has also been considerable progress understanding
certain theories at very strong coupling
\cite{Kovtun:2004de, Bhattacharyya:2008jc, Chesler:2007sv, Kovtun:2005ev},
thanks to the development of gauge/gravity duality
\cite{Maldacena:1997re,Witten:1998qj,GKP}\,.
But very little progress has been made in regimes where a theory is 
both strongly coupled and far from equilibrium.

Heavy ion collisions at the Relativistic Heavy Ion Collider (RHIC)
are believed to produce a deconfined, strongly coupled quark-gluon
plasma (QGP) \cite{Shuryak,Shuryak:2004cy}.
In the initial stages of the collision,
during which the QGP is produced, the system is surely 
far from equilibrium and cannot be described by hydrodynamics.
However, modeling based on near-ideal hydrodynamics
strongly suggests that a hydrodynamic treatment
becomes applicable rather quickly, perhaps on times $\lesssim 1$ fm/c 
after the collision event \cite{Heinz:2004pj}\,.
Understanding the dynamics responsible for such a rapid
approach to local equilibrium from a far-from-equilibrium initial
state is a challenge.

Colliding nuclei at sufficiently high energy
is the only experimentally accessible approach for creating
quark-gluon plasma.
However, analyzing the dynamics --- from the creation of the initial
highly non-equilibrium state,
through partial equilibration, hydrodynamic evolution,
hadronization, and eventual freeze-out ---
via a first principles calculation in QCD is not currently possible.  
Nevertheless,
aspects of this process involving strongly coupled dynamics
can be studied in a controlled setting in a class of theories
which describe non-Abelian plasmas similar to the QGP, and which possess
dual gravitational descriptions.
The best known example is
$\mathcal N = 4$ supersymmetric Yang-Mills (SYM) theory
\cite{Maldacena:1997re}.
In this theory 
one can study the collision
of shock waves
using gauge/gravity duality
\cite{Grumiller:2008va, Gubser:2008pc, AlvarezGaume:2008fx, Lin:2009pn}.
The shock waves have a very small thickness along the collision axis 
and can be localized in the transverse directions \cite{Gubser:2009sx}.
Therefore, qualitatively at least, they resemble the Lorentz-contracted
relativistic nuclei in a heavy ion collision.

In the dual gravitational description,
collisions of shock waves in SYM turn into a problem of 
colliding gravitational shock waves in five dimensions.
The resulting $5d$ numerical relativity problem
is still quite challenging, but may be feasible using
techniques which are adapted from current work in $4d$ numerical relativity.
One purpose of this paper is to begin exploring some of the needed
adaptations, albeit in a setting which is simpler than
colliding shock waves.

The immediate goal of this paper is to study
how quickly a far-from-equilibrium strongly-coupled non-Abelian 
plasma relaxes to a regime in which a hydrodynamic description is accurate.
The answer to this question will necessarily have some sensitivity to how
the initial state is created.
A conceptually simple way to prepare non-equilibrium states
is to start in the ground state,
and then to turn on time-dependent background fields coupled to
operators of interest.  After the background fields are turned off,
one can then watch the subsequent evolution of the system.  
Since a hydrodynamic description requires
that the stress tensor (in the local fluid rest frame)
be nearly isotropic \cite{Arnold:2004ti},
particularly interesting initial states are those in which
the initial stress tensor is driven to be highly anisotropic.
A natural way to do this is to make the spatial geometry in
which the field theory lives be time dependent and anisotropic
\cite{Chesler:2008hg,Bhattacharyya:2009uu}.  

At weak coupling, the addition of energy to the ground state
by a time-dependent gravitational field can be understood in terms of 
particle production.  A time-dependent spacetime geometry will
create quanta \cite{Birrell:1982ix},
and if the time dependence of the deformation in the geometry is anisotropic,
then the momentum distribution of created quanta will be anisotropic as well.  
After the geometry ceases to evolve,
quanta will continue to collide and interact
and eventually
(on a timescale which at weak coupling diverges like $1/\lambda^2$, with
$\lambda$ the `t Hooft coupling)
the system may approach a state in approximate local thermal equilibrium.

This quasi-particle picture breaks down
as the strength of the coupling increases,
and one must understand the process of plasma production and relaxation 
using a different physical description.
For large $\Nc$ SYM, gauge/gravity duality provides an
alternative picture involving black hole formation in five dimensions.
As we discuss in Section~\ref{grav}, 
the gravitational dual will involve a $5d$ curved spacetime with a
$4d$ boundary which has a time dependent geometry.
The boundary geometry
corresponds to the spacetime geometry of the SYM field theory.
A time-dependent deformation in the $4d$ boundary geometry will
produce gravitational radiation which propagates into the fifth
dimension.  This radiation will necessarily produce a black hole
\cite{Chesler:2008hg}.
It is natural that the gravitational description of plasma
formation and relaxation involves horizon formation,
since at late times the system will be in a near-equilibrium
state with non-zero entropy.

The presence of a black hole acts as an absorber of gravitational
radiation and therefore, after the production of gravitational
radiation on the boundary ceases, the $5d$ geometry will relax
onto a smooth and slowly varying form.
This relaxation is dual to the relaxation of
non-hydrodynamic degrees of freedom in the quantum field theory
\cite{Kovtun:2005ev}.  Therefore, by studying the evolution of the
$5d$ black hole geometry, one can gain insight into the
creation and relaxation SYM plasma.

For simplicity, in this paper we limit attention to $4d$ geometries
which have two dimensional spatial homogeneity and $O(2)$ rotation invariance
in the $\bm x_{\rm \perp} \equiv \{x^1,\ x^2\}$ directions, and which
are invariant under boosts in the $x_\| \equiv x^3$ direction.
As we discuss in Section~\ref{grav}, this reduces the gravitational
dynamics to a system of two-dimensional PDEs, which we solve
numerically.  Besides making the gravitational calculation simpler,
these assumptions serve an additional purpose.
With these symmetries, the late time asymptotics of the $5d$ geometry
(and the corresponding asymptotics of the stress tensor)
are known analytically \cite{Janik:2005zt,Kinoshita:2008dq,Heller:2009zz}.
We will therefore be able to compare directly our numerical results,
valid at all times, to the known late time asymptotics.

Boost invariance implies that the natural coordinates to use are
proper time $\tau$ and rapidity $y$
(with $x^0 \equiv \tau \, \cosh y$ and $x_\| \equiv \tau \,\sinh y$).
In these coordinates, the metric of $4d$ Minkowski space
(in the interior of the $\tau = 0$ cone) is
$
    ds^2 = -d\tau^2 +  d \bm x_{\perp}^2 +  \tau^2 \, dy^2
$.
A deformation of the geometry,
respecting the above symmetry constraints,
induced by a time-dependent shear may be written in the form
\begin{equation}
\label{boundarygeometry}
    ds^2 = -d\tau^2
    +  e^{\gamma(\tau)} \, d \bm x_{\perp}^2
    +  \tau^2 \, e^{-2 \gamma(\tau)} \, dy^2 \,.
\end{equation}
The function $\gamma(\tau)$ characterizes the time-dependent shear;
neglecting $4d$ gravity, $\gamma(\tau)$ is a function
one is free to choose arbitrarily.  
For this study, we chose
\begin{align}
\label{beta}
    \gamma(\tau) =
    c\,
    &\Theta \left (1 - {(\tau {-} \tau_0)^2}/{\Delta^2} \right )
    \left [ 1 - {(\tau {-} \tau_0)^2}/{\Delta^2} \right]^6
\nonumber\\ & {} \times
    e^{  -1/\left [1 - {(\tau {-} \tau_0)^2}/{\Delta^2} \right ] },
\end{align}
with $\Theta$ the unit step function.
(Inclusion of the $ [ 1 - {(\tau {-} \tau_0)^2}/{\Delta^2} ]^6$
factor makes the first few derivatives of $\gamma(\t)$
better behaved as $\t{-}\tau_0 \to \pm \Delta$.)
The function $\gamma(\t)$ has compact support and is infinitely
differentiable;
$\gamma(\t)$ and all its derivatives vanish at the endpoints
of the interval $(\tau_i, \tau_f)$,
with $\tau_i \equiv \tau_0 -\Delta $ and $\tau_f \equiv \tau_0 + \Delta$.
We choose $\tau_0 \equiv {\textstyle \frac{5}{4} } \Delta$ so the geometry
is flat at $\tau = 0$.%
  \footnote
    {%
    Choosing $\tau_0 \ge \Delta$ is convenient for numerics as
    our coordinate system becomes singular on the $\tau = 0$ lightcone.
    The particular choice $\tau_0 = \frac 54 \Delta$ was made
    so that our numerical results (which begin at $\tau = 0$)
    contain a small interval of unmodified geometry before the
    deformation turns on.  For an interesting discussion of non-equilibrium 
    boost invariant states near $\tau = 0$ see Ref.~\cite{janikandheler}.
    }
We choose to measure all dimensionful quantities in units where $\Delta = 1$
(so $\tau_i = 1/4$ and $\tau_f = 9/4$).

\begin{figure}[t]
\includegraphics[scale=0.35]{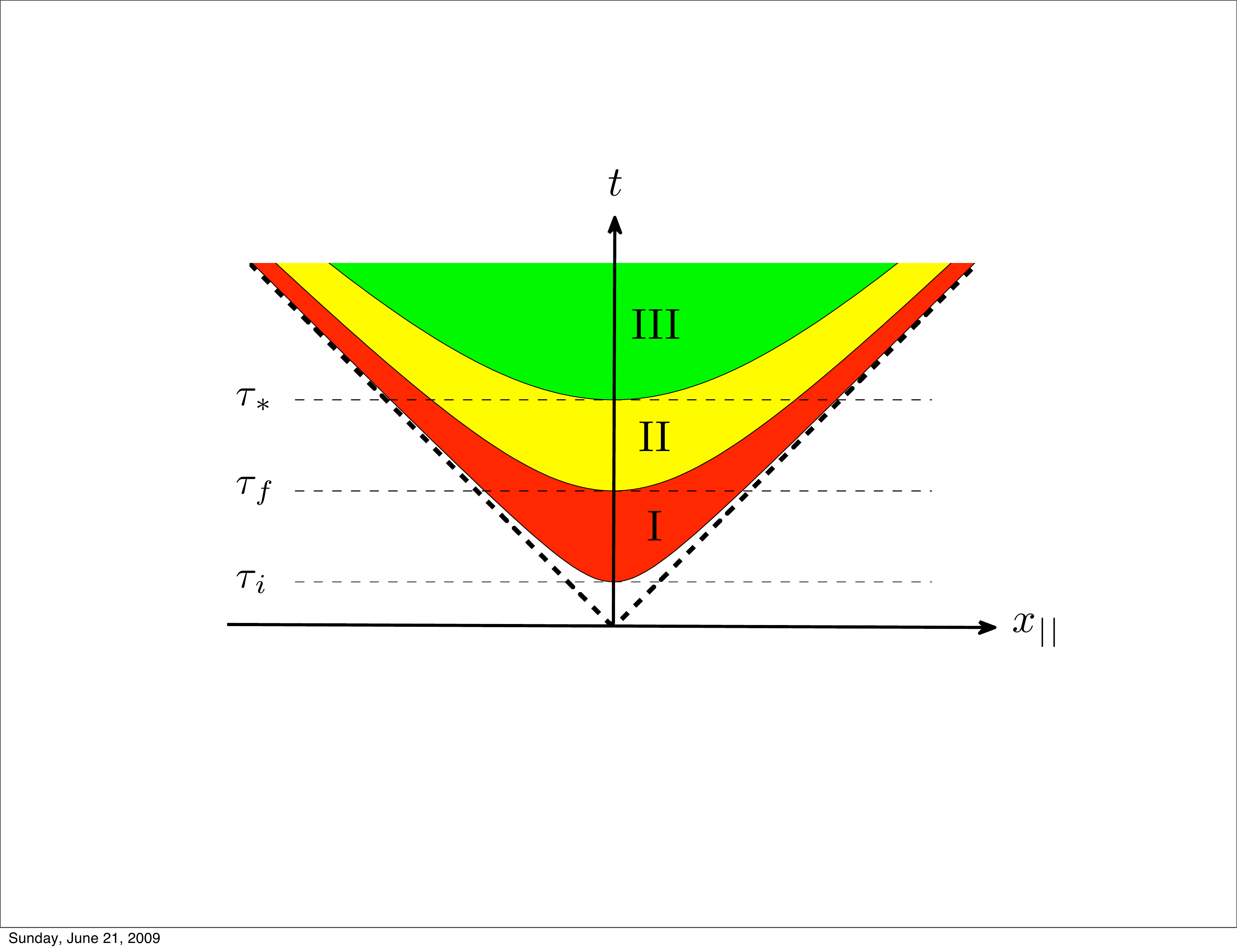}
\caption
  {
  \label{spacetime}
  A spacetime diagram depicting several stages of the evolution of the 
  field theory state in response to the changing spatial geometry.  
  At proper time $\tau = \tau_i$,
  the $4d$ spacetime geometry starts to deform.
  The region of spacetime where the geometry undergoes time-dependent
  deformation is shown as the red region, labeled I.  
  After proper time $\tau = \tau_f$,
  the deformation in $4d$ spacetime geometry turns off
  and the field theory state is out of equilibrium.
  From proper time $\tau_f$ to $\tau_*$,
  shown as the yellow region labeled II,
  the system is significantly anisotropic and
  not yet close to local equilibrium.
  After time $\tau_*$, shown in green and labeled III,
  the system is close to local equilibrium and the evolution
  of the stress tensor is well-described by hydrodynamics.
  }
\end{figure}

Fig.~\ref{spacetime} shows a spacetime diagram 
schematically depicting several stages in the evolution of the SYM state.
Hyperbola inside the forward lightcone are constant $\tau$ surfaces.
Prior to $\tau = \tau_i$, the system is in the ground state.
The region of spacetime where the geometry is deformed from flat space is 
shown as the red region labeled I in Fig.~\ref{spacetime}.
At coordinate time $t = \tau_i$ the geometry of spacetime
begins to deform in the vicinity of $x_{\|} = 0$.  As time progresses,
the deformation splits into two localized regions centered about
$x_{\|} \sim \pm t$, which subsequently separate and
move in the $\pm x_{\|}$ directions at the speeds asymptotically
approaching the speed of light.  
After the ``pulse'' of spacetime deformation passes,
the system will be left in an excited, anisotropic,
non-equilibrium state.  That is, the 
deformation in the geometry will have
done work on the field theory state.
This region, labeled II, is shown in yellow in Fig.~\ref{spacetime}.
It is in this region that we can study the relaxation of a
far-from-equilibrium non-equilibrium state.
After some amount of proper time $\tau_*$,
the system will have relaxed to a point where
a hydrodynamic description of the continuing evolution
is accurate.
This final hydrodynamic regime is
shown schematically in green, and labeled III,
in Fig.~\ref{spacetime}.
As the late time hydrodynamic solution to boost invariant flow is known
analytically, we choose to define $\tau_*$ precisely as the time beyond
which the stress tensor coincides with the hydrodynamic approximation to
within $10\%$.

Our task then is to find $\tau_*$ and in particular, see
how it correlates with quantities such as the effective temperature $T_*$
at time $\tau_*$.
In the $c \rightarrow \infty$ limit,
which corresponds to a diverging size of
the deformation in the $4d$ geometry,
it is inevitable that $\tau_* $ approaches $\tau_f$.
This is natural in conformal theories,
since relaxation times of non-hydrodynamic degrees of freedom
are set by the local energy density,
and this diverges when $c\rightarrow \infty$.
In other words,
in the limit where $1/T_* \ll \Delta$,
the system responds adiabatically to the 
deformation in the geometry and non-hydrodynamic degrees of
freedom can remain in equilibrium.
A hydrodynamic description (without driving terms)
will be accurate the moment the geometry stops changing.
Hence, in this limit one learns nothing about the dynamics
associated with the relaxation of non-hydrodynamic modes.

More interesting is the case where the effective temperature
satisfies $1/T_* \gtrsim \Delta$.
This is the regime we will study.
Within this regime, the system can be significantly out-of-equilibrium
after the $4d$ geometry becomes flat.
When this is the case,
we find that the entire process of plasma creation
and relaxation to approximate local equilibrium
({\em i.e.}, to a point where subsequent evolution is accurately
described by viscous hydrodynamics)
occurs over a time which varies between one and two times $1/T_*$.

This result is consistent with the findings in our earlier work
\cite {Chesler:2008hg}
where we studied isotropization in a
homogeneous strongly coupled $\Nfour$ SYM plasma.
In that work, all spatial gradients vanished.
There was no
excitation whatsoever of hydrodynamic degrees of freedom,
and the system relaxed exponentially toward equilibrium.
In contrast, the dynamics of the boost-invariant plasma in
the present work involves both hydrodynamic and non-hydrodynamic
degrees of freedom.
The results we present display a rather clear separation
between far-from-equilibrium response,
which cannot be described by hydrodynamics,
followed by later ``near local equilibrium'' dynamics which is
accurately described by viscous hydrodynamics.
A noteworthy finding is that the domain of utility of hydrodynamics
is not limited by when higher order terms in the hydrodynamic
expansion become comparable to the lowest order viscous terms,
rather it is determined by the relative
importance of non-hydrodynamic degrees of freedom.

\section{{Gravitational description}}
\label{grav}

Gauge/gravity duality \cite{Maldacena:1997re}  provides a
gravitational description of large $\Nc$ SYM in which the $5d$
dual geometry is governed by Einstein's equations 
with a cosmological constant.
Einstein's equations imply that the boundary metric
$g_{\mu\nu}^{\rm B}(x)$,
which characterizes the geometry of the spacetime boundary,
is dynamically unconstrained.  
The specification of $g_{\mu\nu}^{\rm B}(x)$ serves as a boundary condition 
for the $5d$ Einstein equations.
This reflects the fact that
$4d$ gravitational dynamics is neglected;
the dual field theory residing on the boundary
responds to the boundary geometry but
does not back-react on the $4d$ boundary geometry.

Diffeomorphism and spatial $3d$ translation invariance,
together with our assumed $O(2)$ rotation invariance,
allows one to chose a $5d$ bulk metric of the form
\begin{align}
\label{metric}
ds^2 = &-A \, d\t^2 +
\Sigma^2 \big [ e^{B} d \bm x_{\perp}^2 + e^{-2 B} dy^2 \big ] + 2 dr \,d\t\,,
\end{align}
where
$A$, $B$, and $\Sigma$ are all functions of the bulk radial coordinate $r$
and time $\t$ only.
The coordinates $\t$ and $r$ are generalized
infalling Eddington-Finkelstein coordinates.
Infalling radial null geodesics have constant values of $\t$
(as well as $\bm x_\perp$ and $y$).
Outgoing radial null geodesics satisfy ${d r}/{d\t } = \frac{1}{2}A$.
The geometry in the bulk at $\t>0$ corresponds to the causal future
of $\tau=0$ on the boundary.
The form of the metric (\ref{metric}) is invariant 
under the residual diffeomorphism  
$
    r \rightarrow r + f(\t),
$
where $f(\t)$ is an arbitrary function.

With a metric of the form (\ref{metric}),
Einstein's equations may be written very compactly as
\begin{subequations}
\begin{eqnarray}
\label{Seq}
0 &=& \Sigma \, (\dot \Sigma)' + 2 \Sigma' \, \dot \Sigma - 2 \Sigma^2\,,
\\ \label{Beq}
0 &=& \Sigma \, (\dot B)' + {\textstyle \frac{3}{2}}
    \big ( \Sigma' \dot B + B' \, \dot \Sigma \big )\,,
\\  \label{Aeq}
0 &=& A'' + 3 B' \dot B - 12 \Sigma' \, \dot \Sigma/\Sigma^2 + 4\,,
\\  \label{Cr}
0 &= & \ddot \Sigma
    + {\textstyle \frac{1}{2}} \big( \dot B^2 \, \Sigma - A' \, \dot \Sigma \big)\,,
\\ \label{Cv}
0 &=& \Sigma'' + {\textstyle \frac{1}{2}} B'^2 \, \Sigma\,,
\end{eqnarray}
\label{Eeqns}%
\end{subequations}
where, for any function $h(r,\t)$,
\begin{equation}
    h' \equiv \partial_r h, \qquad
    \dot h \equiv \partial_\t h + {\textstyle \frac{1}{2}} A \, \partial_r h\,.
\end{equation}
The derivative $h'$ is a directional derivative of $h$ along infalling radial
null geodesics, while
the derivative $\dot h$ is the directional derivative of $h$
along outgoing null radial geodesics.
Eqs.~(\ref{Cr}) and (\ref{Cv}) are constraint equations;
the radial derivative of Eq.~(\ref{Cr})
and the time derivative of Eq.~(\ref{Cv})
are implied by Eqs.~(\ref{Seq})--(\ref{Aeq}).

The above set of differential equations must be solved subject to
boundary conditions imposed at $r = \infty$.
The requisite condition is simply that the boundary metric
$g_{\mu\nu}^{\rm B}(x)$
coincide with our choice (\ref{boundarygeometry}) of the $4d$ geometry.
In particular, we must have
\begin{subequations}%
\label{bc}%
\begin{align}
&\lim_{r \rightarrow \infty} \Sigma(r,\t)/r \equiv \t^{1/3}\,,
\\
&\lim_{r \rightarrow \infty} B(r,\t) \equiv -{\textstyle \frac{2}{3}} \ln \t + \gamma(\t)\,.
\end{align}
\end{subequations}
One may fix the residual diffeomorphism invariance 
by also demanding that
\begin{equation}
\label{gaugefix}
\lim_{r \rightarrow \infty} \left [ A(r,\t) - r^2 \right ]/r = 0\,.
\end{equation}
These boundary conditions,
plus initial data satisfying the constraint (\ref{Cv}) 
on some $\t = \tau_i$ slice,
uniquely specify the subsequent evolution of the geometry.

Near the boundary one may solve
Einstein's equations with a power series expansion in $r$.  
Specifically, $A$, $B$ and $\Sigma$ have asymptotic expansions of the form
\begin{subequations}
\label{series}
\begin{eqnarray}
A(r,\t) &= &  \sum_{n=0} \left [\, a_{n}(\t) + \alpha_{n}(\t) \log r \right ] r^{2-n}\,,
\\
B(r,\t) &= & \sum_{n=0} \left [\, b_{n}(\t) + \beta_{n}(\t) \log r \right ] r^{-n}\,,
\\
\Sigma(r,\t) &= & \sum_{n=0} \left [\, s_{n}(\t) + \sigma_{n}(\t) \log r \right ] r^{1-n}\,.
\label{eq:Sexp}
\end{eqnarray}
\end{subequations}
The boundary conditions (\ref{bc}) and (\ref{gaugefix}) imply that
$b_{0}(\t) \equiv -{\textstyle \frac{2}{3}} \ln \t + \gamma(\t)$,
$s_{0}(\t) \equiv \t^{1/3}$,
$a_{0}(\t) \equiv 1$,
$a_{1}(\t) \equiv~0$,
and that the coefficients of the corresponding logarithmic terms vanish.
Substituting the above expansions into Einstein's equations
and solving the resulting equations order by order in $r$,
one finds that there is one undetermined coefficient, $b_4(\t)$.
All other coefficients are determined by the boundary geometry,
Einstein's equations, and $b_4(\t)$.%
\footnote
    {%
    The coefficient $a_4$ is determined by a first order ordinary
    differential equation, which can be obtained
    from the condition that the SYM stress tensor be covariantly conserved.
    All other coefficients are determined algebraically from
    $b_{0}(\t)$, $b_{4}(\t)$, $a_4(\t)$ and their derivatives. 
    }

Given a solution to Einstein's equations, the SYM stress tensor is determined
by the near-boundary behavior of the $5d$ metric \cite{deHaro:2000xn}\,.
If $S_{\rm G}$ denotes the gravitational action, then
the SYM stress tensor is given by
\begin{equation}
    T^{\mu \nu}(x) = \frac{2}{\sqrt{-g^{\rm B}(x)}} \>
    \frac{\delta S_{\rm G}}{\delta g^{\rm B}_{\mu \nu}(x)}\,.
\end{equation}
By substituting the above series expansions into the variation
of the on-shell gravitational action, one may compute the expectation
value of the stress tensor in terms of the expansion coefficients.  This 
procedure has been carried out in Ref.~\cite{deHaro:2000xn}, so we simply quote
the results.  In terms of the expansion coefficients,
the SYM stress tensor reads
\begin{equation}
T^{\mu}_{\ \nu} = \frac{N_c^2}{2 \pi^2} \>
{\rm diag} (-\mathcal E,\mathcal P_{\perp},\mathcal P_{\perp},\mathcal P_{||} ) \,,
\end{equation}
where
\begin{subequations}
\label{stress}
\begin{align}
\mathcal E = {}&   -{\textstyle \frac{3}{4} } a_4 + \widetilde{\mathcal E},
\\
 \mathcal P_{\perp} ={}& -{\textstyle \frac{1}{4} } a_4 +b_4+  \widetilde{\mathcal P}_{\perp},
\\ 
 \mathcal P_{||}  ={}&  - {\textstyle \frac{1}{4} } a_4 -2 b_4 +  \widetilde{\mathcal P}_{||},
\end{align}
\end{subequations}
and
\begin{subequations}
\begin{align}
    \widetilde{\mathcal E}
    \equiv&
    -\tfrac{5}{288} \, \gamma_1 \,{\t^{-3}} 
    + \tfrac{5}{1152} ( 21 \gamma_1^2 + 4 \gamma_2 )\,{\t^{-2}}
\nonumber
\\ &
    - \tfrac{1}{96} (3\gamma_1^3 - 8 \gamma_1 \gamma_2 - \gamma_3)\,{\t^{-1}}
\nonumber
\\ &
    +\tfrac{1}{256} (3\gamma_1^4 + 14 \gamma_2^2 - 4 \gamma_1  \gamma_3)\,,
\\[5pt]
    \widetilde{\mathcal P}_{\perp}
    \equiv& 
    -\tfrac{1}{6} \,\t^{-4}
    + \tfrac{227}{288}\, \gamma_1\, \t^{-3}
    - \tfrac{1}{3456}(2397 \gamma_1^2 + 1444 \gamma_2)\,\t^{-2}
\nonumber
\\ & 
    +\tfrac{1}{576}(57 \gamma_1^3 + 488 \gamma_1  \gamma_2 + 70  \gamma_3)\,\t^{-1}
\nonumber
\\ & 
    +\tfrac{1}{768}(21  \gamma_1^4 - 468  \gamma_1^2  \gamma_2 + 10  \gamma_2^2 +4  \gamma_1  \gamma_3 + 64  \gamma_4)\,,
\nonumber
\\
\\
    \widetilde{\mathcal P}_{||}
    \equiv& 
    \tfrac{1}{3}\, \t^{-4}
    - \tfrac{449}{288}\, \gamma_1\, \t^{-3}
    + \tfrac{1}{3456}(5379  \gamma_1^2 + 2828  \gamma_2)\,\t^{-2}
\nonumber
\\ &
    - \tfrac{1}{288}(120  \gamma_1^3 + 458  \gamma_1  \gamma_2 + 73  \gamma_3)\,\t^{-1}
\nonumber
\\ &
    +\tfrac{1}{768}(21\gamma_1^4 + 936 \gamma_1^2  \gamma_2 + 10  \gamma_2^2 + 4  \gamma_1  \gamma_3-128  \gamma_4)\,,
\nonumber
\\
\end{align}
\end{subequations}
with $\gamma_n \equiv d^n \gamma/d\t^n$.

\section{{Numerics}}

One may solve the Einstein equations (\ref{Seq})--(\ref{Aeq}) for
the time derivatives $\dot\Sigma$, $\dot B$, and $A''$.
Define
\begin{subequations}
\begin{align}
\label{Theta}
    \Theta(r,\t) \equiv
    \int_r^{\infty} &dw \left [\Sigma(w,\t)^3 - h_1(w,\t) \right ] - H_1(r,\t)\,,
\\
    \Phi(r,\t) \equiv 
    \int_r^{\infty} &dw \left [2 \Theta(w,\t) B'(w,\t)\, \Sigma(w,\t)^{-3/2}\right.
\nonumber \\ &\quad{}
    - h_2(w,\t)  \Bigr ] - H_2(r,\t)\,,
\label{Phi}
\end{align}
\label{ThetaPhi}%
\end{subequations}
where $H_n$ is an indefinite radial integral of $h_n$,
\begin{equation}
\label{heq}
h_n = H'_n \,.
\end{equation}
Then Eqs.~(\ref{Seq})--(\ref{Aeq}) are solved by
\begin{subequations}
\label{reducedeqns}
\begin{eqnarray}
\label{Sigmadot}
\dot \Sigma &=& -2 \Theta \, \Sigma^{-2}, 
\\ \label{Bdot}
\dot B &=& -\coeff {3}{2} \,\Phi \Sigma^{-3/2}\,,
\\ \label{Aeq2}
A'' &=&
- 4 -24  \Theta \,  \Sigma' \Sigma^{-4} + \coeff 92 \Phi B' \, \Sigma^{-3/2}  \,.
\end{eqnarray}
\end{subequations}
The functions $h_n(r,\t)$ are not constrained by Einstein's equations ---
their presence inside the integrands of Eq.~(\ref{ThetaPhi})
are compensated by the subtraction of their integrals $H_n(r,\t)$.
However, 
since we have chosen the upper limit of integration in Eq.~(\ref{ThetaPhi})
to be $r = \infty$,
the functions $h_n(r,\t)$ must be suitably
chosen so that the integrals (\ref{ThetaPhi}) are convergent.
The simplest choice to accomplish this is to set $h_1(r,\t)$
equal to the asymptotic expansion of $\Sigma(r,\t)^3$ up to order $1/r^k$,
for some $k > 1$,
and to set $h_2(r,\t)$ equal to the asymptotic expansion
of $2 \Theta(r,\t)B'(r,\t)/\Sigma(r,\t)^{3/2}$ up to order $1/r^k$.
In our numerical solutions reported below, we use $k \geq 4$.
This choice makes the large $r$ contribution to the integrals in
Eq.~(\ref{ThetaPhi}) quite small and consequently reduces cutoff dependence.
As the coefficients of the series expansions (\ref{series}) only depend on
$b_{0}(\t)$ and $b_{4}(\t)$ and their $\t$ derivatives,
this choice determines $h_n(r,\t)$ in terms of one
unknown function $b_{4}(\t)$. 

With the subtraction functions $h_n$ specified by the aforementioned asymptotic 
expansions, integrating Eq. (\ref{heq}) fixes
the compensating integrals $H_n$ 
up to an integration constant which is an arbitrary function of $\t$.
Integrating Eq.~(\ref{Aeq2}) for $A(r,\t)$ introduces two further
($\t$ dependent) constants of integration.
The most direct route for fixing these constants of integration is to 
match the large $r$ behavior of the expressions (\ref{Sigmadot}) and (\ref{Bdot})
and the integrated version of Eq.~(\ref{Aeq2}) to the
corresponding expressions obtained from the series expansions (\ref{series}).
This fixes all integration constants in terms of $b_0$ and~$b_4$.

Our algorithm for solving the initial
value problem with time dependent boundary conditions is as follows.
At time $\tau_i$ the geometry is AdS$_5$ with the metric
\begin{align}
\label{ads}
    ds^2 = r^2
    \left [ {-} d\t^2 + d \bm x_{\perp}^2 + \big( \t + \tfrac {1}{r} \big)^2 d y^2 \right ]
+ 2 dr \, d\t \,.
\end{align}
Therefore, at the initial time $\tau_i$ we have
\begin{subequations}
\label{initialgeometry}
\begin{eqnarray}
B(r,\tau_i) &=&  -{\tfrac{2}{3} \ln \big(\tau_i + \tfrac{1}{r} \big) }\,,
\\
\Sigma(r,\tau_i) &=& r \big (\tau_i + {\textstyle \frac{1}{r} }\big )^{1/3}\,, 
\\[3pt]
A(r,\tau_i) &=& r^2\,.
\end{eqnarray}
\end{subequations}
With $A(r,\tau_{i})$,
$B(r,\tau_{i})$ and $\Sigma(r,\tau_{i})$ known, one can then compute 
the time derivatives $\partial_\t B(r,\tau_{i})$
and $\partial_\t \Sigma(r,\tau_{i})$
from Eqs.~(\ref{Bdot}) and (\ref{Sigmadot}),
and step forward in time,
\begin{eqnarray}
    B(r,\tau_{i} + \Delta \tau) &\approx& B(r,\tau_{i}) + \partial_\t B(r,\tau_{i})
    \Delta \tau \,,
    \\
    \Sigma(r,\tau_{i} + \Delta \tau) &\approx& \Sigma(r,\tau_{i}) + \partial_\t \Sigma(r,\tau_{i}) \,
    \Delta \tau \,.
\end{eqnarray}
With $ B(r,\tau_{i} {+} \Delta \tau)$ and
$\Sigma(r,\tau_{i} {+} \Delta \tau)$ known, one can then
integrate Eq.~(\ref{Aeq2}) to determine $A(r,\tau_{i} {+} \Delta \tau)$.
With the complete geometry on the time slice
$\tau = \tau_i {+} \Delta \tau$ determined,
one may then repeat the entire process
and take another step forward in time.%
\footnote
    {%
    Because we are working with a discretized version of Einstein's equations,
    the discretized version of the constraint equation (\ref{Seq}) is not
    automatically implied by the discretized version of the other Einstein
    equations.
    To minimize the amount of accumulated error, we also monitor the
    accuracy of the constraint equation (\ref{Seq}),
    and make tiny adjustments to $\Sigma$ to prevent growing
    violation of the constraint.
    } 

An important practical matter is fixing the computation domain in $r$ ---
how far into the bulk does one want to compute the geometry?
If a horizon forms, then one may excise the geometry inside the horizon
as this region is causally disconnected
from the geometry outside the horizon.
Furthermore, one must excise the geometry to avoid
singularities behind horizons \cite{Anninos:1994dj}\,.  
To perform the excision,
one first identifies the location of an
apparent horizon (an outermost marginally trapped surface)
which, if it exists, must lie inside an event horizon \cite{Wald:1984rg}\,.
We have chosen to make the incision slightly inside the location
of the apparent horizon.
For the metric (\ref{metric}), the location $r_h(\t)$ of the apparent horizon
is given by the outermost point where
$\dot \Sigma(r_h(\t),\t) = 0$ or, from Eq.~(\ref{Sigmadot}),
$
\Theta(r_h(\t),\t) = 0\,.
$

\section{{Results and Discussion}}
\label{results}

\begin{figure}[t]
\includegraphics[scale=0.35]{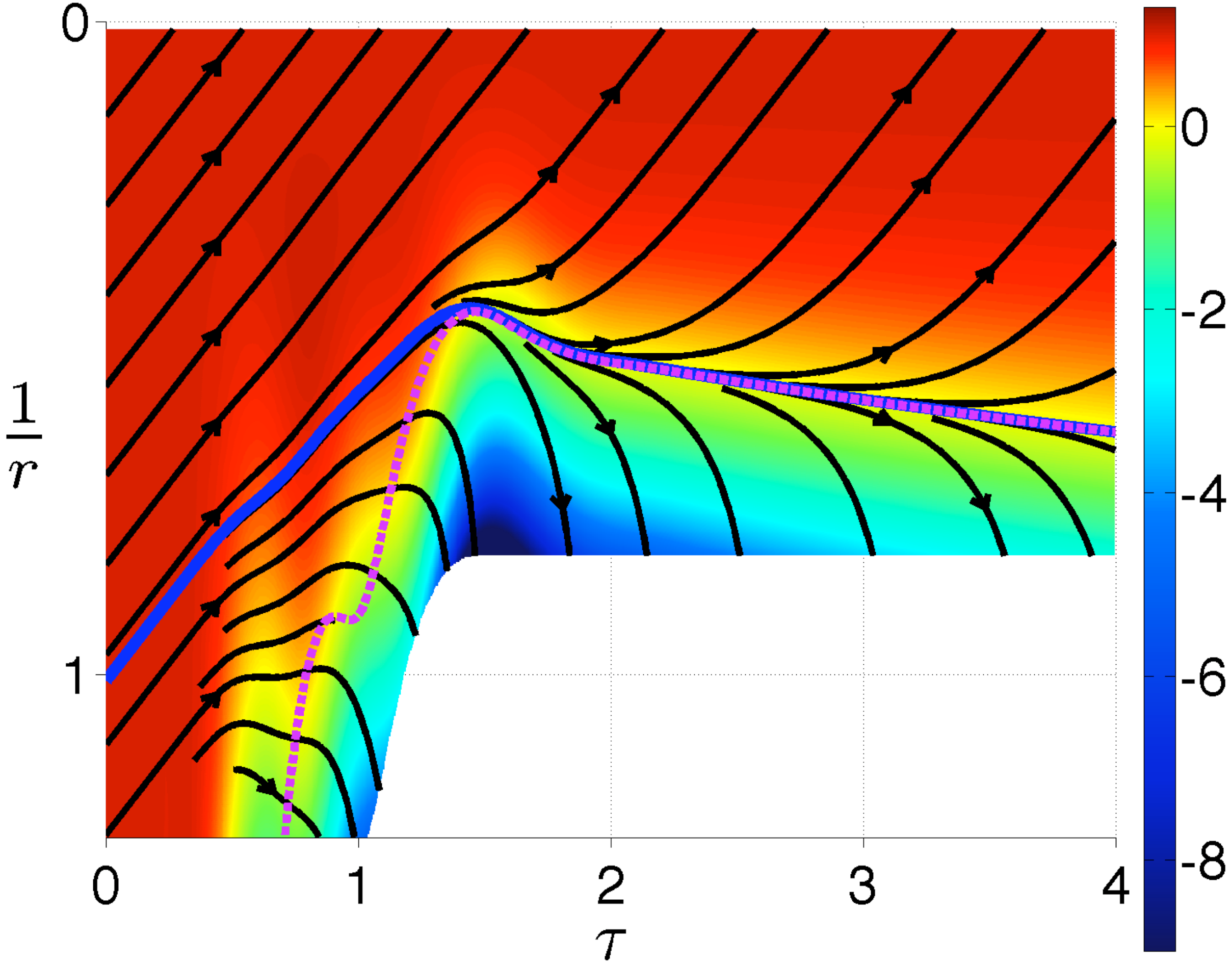}
\caption
  {
  \label{geodesics}
  The congruence of outgoing radial null geodesics.  The surface coloring
  displays $A/r^2$.  Before time $\tau_i = 1/4$ this quantity equals one.
  The excised region lies inside the apparent horizon,
  which is shown by the dashed magenta line.
  The geodesic shown as a solid blue line is the event horizon;
  it separates geodesics which escape to the boundary from those which cannot
  escape.
  }
\end{figure}

We first discuss our results from the $5d$ gravitational perspective and 
present data for $c=1$.  
Results for other values of $c$ are presented below,
but the qualitative features of the results
are independent of the value of $c$.
Fig.~\ref{geodesics} shows a congruence of outgoing radial null
geodesics for $c = 1$.  The geodesics are obtained by integrating
${dr}/{d\t} = \half A(r,\t)$.
The colored surface in the plot displays the value of $A/r^2$.
Excised from the plot is a region of the geometry 
behind the apparent horizon, whose location is
shown by the magenta dotted line.

At times $\tau < \tau_i = 1/4$,
the boundary geometry is static and $A/r^2 = 1$.
The outgoing geodesic congruence at early times therefore satisfies
\begin{equation}
\label{lincongr}
\t + 2/r = \mathrm{const.}\,,
\end{equation}
and hence appears as parallel straight lines on the left side
of Fig.~\ref{geodesics}.
These are just radial geodesics in AdS$_5$, which is the geometry
dual to the initial zero temperature ground state.
After time $\tau_i$
the boundary geometry starts to change,
$A/r^2$ deviates from unity, and the congruence 
departs from the zero temperature form (\ref{lincongr}).  

Perhaps the most dramatic feature in Fig.~\ref{geodesics}
is the formation of a bifurcation in the congruence of geodesics.  As is evident
from Fig.~\ref{geodesics}, at late times
some geodesics escape up to the boundary and some plunge 
deep into the bulk.  Separating escaping from plunging geodesics is
precisely one geodesic that does neither.
This geodesic, shown as the solid blue curve in the figure, 
defines the location of a null surface beyond which all events 
are causally disconnected from observers on the boundary.  This surface 
is the event horizon of the geometry.

After the time $\tau_f = 2.25$,
the boundary geometry becomes flat and unchanging,
no additional gravitational radiation is produced,
and the bulk geometry approaches a slowly evolving form.  
The rapid relaxation of high frequency modes can clearly be seen
in the behavior of $A/r^2$ shown in Fig.~\ref{geodesics} ---
all of the high frequency structure in the plot appears only during
the time interval where the boundary geometry is changing and creating 
gravitational radiation.
Physically, the rapid relaxation of high frequency modes occurs because
the horizon acts as an absorber of gravitational radiation and 
low frequency modes simply take more time to  
fall into the horizon than high frequency modes.
Therefore, as time progresses the geometry
relaxes onto a smooth universal form whose temporal variations 
become slower and slower as $\tau \rightarrow \infty$.  

One can systematically construct late-time asymptotic expansions
of boost-invariant solutions to Einstein's equations
\cite{Bhattacharyya:2008jc}.
The expansion, which is a power series expansion in gradients,
is dual to the hydrodynamic expansion in the field theory.
This is natural, as the late time evolution of the field theory
state in conformal $\Nfour$ SYM must be described by hydrodynamics.
In the gravitational setting,
the metric is expanded in terms of $4d$ spacetime gradients of
slowly varying fields.
For the case of boost invariant flow, each spacetime derivative introduces
a factor of $1/(\Lambda \tau)^{2/3}$ into the solution,
where $\Lambda$ is an energy scale which characterizes the
initial energy density \cite{Janik:2005zt}.
The numerical coefficients of the expansion are related to
transport coefficients in the dual gauge theory,
and are independent of the initial conditions used
to create the black hole geometry.
Therefore, at asymptotically late times all sensitivity to the details of
the initial conditions used to created the black hole geometry is
isolated within the energy scale $\Lambda$, up to exponentially
decreasing corrections to the late time behavior.

At asymptotically late times,
the boost invariant gradient expansion
of Ref.~\cite{Janik:2005zt} yields a metric 
\begin{align}
\label{boostblackbrane}
ds^2 = r^2 \bigg [ {-} \Big (1 {-} \frac{r_h^4 }{ r^4} \Big ) d\t^2 + d \bm x_{\perp}^2 + \t^2d y^2 \bigg ]
+ 2 dr d \t,
\end{align}
where $r_h(\t) \approx \pi \Lambda /(\Lambda \t)^{1/3}$
is the approximate location
of the event and apparent horizons, whose positions
asymptotically coincide at late times.
The asymptotic metric (\ref{boostblackbrane}) has a Hawking temperature
\begin{equation}
\label{hawking}
T_{\rm Hawking} = \Lambda/(\Lambda \t)^{1/3},
\end{equation}
which is proportional to the horizon radius $r_h(\t)$.
As time progresses, the horizon slowly falls deeper into the bulk,
and the temperature of the black hole decreases as $\t^{-1/3}$.
The falling of the horizon into the bulk, as an inverse power of $\t$,
is clearly visible in the
numerical data presented in Fig.~\ref{geodesics}.

\begin{figure}[t]
\includegraphics[scale=0.35]{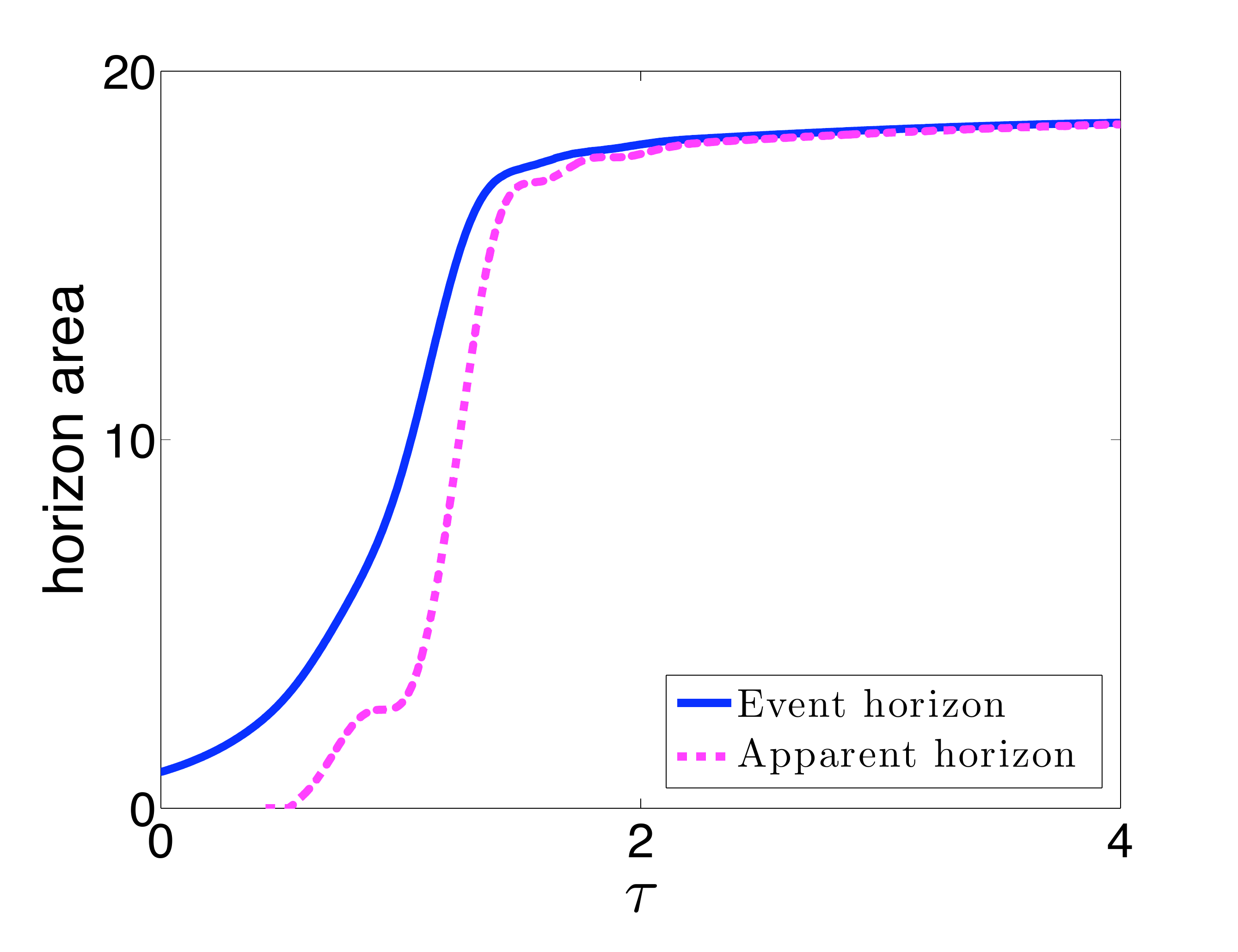}
\vspace*{-5pt}
\caption
    {\label{area}
    Area of the event horizon and apparent horizon,
    per unit rapidity,
    as a function of proper time $\tau$.
    The growth of the apparent horizon area,
    shown by the magenta dotted line, is causally
    connected to the changing boundary geometry.  In contrast, the growth
    of the event horizon area, shown as a solid blue line,
    is non-zero at arbitrarily early times,
    long before the boundary geometry has started to change.
    }
\end{figure}

Fig.~\ref{area} shows a plot of the area (per unit rapidity)
of the event and apparent horizons,
again for $c = 1$, as a function of $\t$.  
The area (per unit rapidity) of the apparent horizon
is given by $\Sigma(r_h(\tau),\tau)^3$
where $r_h(\tau)$ is the apparent horizon location (given by a zero
of $\dot\Sigma$).
The area (per unit rapidity) of the event horizon is also given by
$\Sigma^3$, but instead evaluated on the null geodesic defining the
event horizon.
The area of the apparent horizon starts off at zero,
and grows rapidly for $\tau$ in the interval $(\tau_i,\tau_f)$.
This is to be expected, as it is during this interval of time
that the rapid variation of the boundary geometry produces infalling
gravitational radiation which is subsequently absorbed by the horizon.
As radiation is absorbed, the horizon area must grow. 
After the production of radiation ceases, the 
the geometry relaxes onto the asymptotic form (\ref{boostblackbrane})
and the area (per unit rapidity)
of the apparent and event horizons slowly approach a constant.
{}From the figure, one sees that the growth of the 
apparent horizon area changes rather abruptly near time $\tau_f$.
This reflects of the fact that the boundary geometry 
ceases to produce infalling radiation after time $\tau_f$.
The flux of radiation through the horizon decreases
dramatically after $\tau_f$ and correspondingly,
so does the growth of the apparent horizon area.  

In contrast to the apparent horizon area, which is non-zero only
in the causal future of the boundary time $\tau_i$,
the event horizon area is non-zero arbitrarily far in the past,
long before the boundary geometry starts to change.
This reflects the teleological nature of event horizons.
The event horizon separates events which are causally disconnected 
from boundary observers.  As Fig.~\ref{geodesics} clearly shows,
even before the boundary geometry has started to change there are events
which are causally disconnected from the boundary.
These events are, by definition, behind the event horizon.
Simply put, the black hole exists before the boundary
deformation has begun!

\begin{figure}[t]
\includegraphics[scale=0.35]{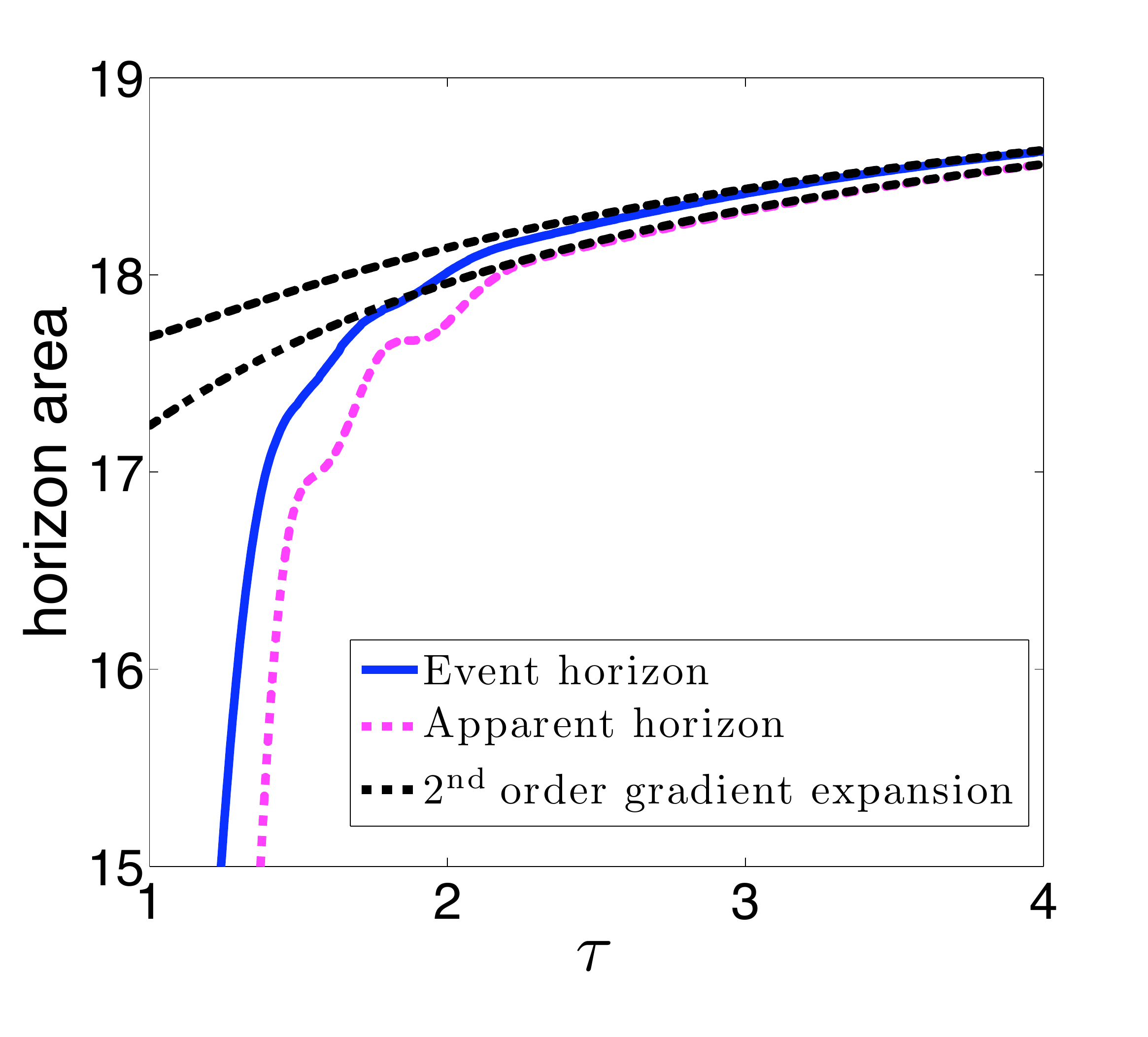}
\caption
    {\label{areagrad}
    Close-up view of the
    event horizon and apparent horizon areas,
    per unit rapidity,
    as a function of proper time $\tau$,
    together with their corresponding asymptotic expressions
    (\ref{asmarea}).
    Both horizon areas are very well approximated by
    their asymptotic expansions, at second order in gradients,
    after time $\tau_f = 2.25$
    when the boundary geometry becomes flat.
    Note the rather abrupt change in the growth
    of the apparent horizon area at~$\tau_f$.
    }
\end{figure}

Because the radial geodesic defining the event horizon is moving outwards
at the speed of light,
before the boundary geometry starts to change
the area of the event horizon grows like $4 (k + \tau)/(k - \tau)^{3}$,
where $k$ is the value of $\tau + 2/r$ on the geodesic
defining the event horizon.
The appropriate value of the constant $k$ 
can only be determined when the entire future
of the geometry is known.%
\footnote
  {%
  This manifests itself as follows.
  At asymptotically late times, the location of the event horizon 
  coincides with the zero of $A(r,\tau)$, so
  the unique outgoing radial geodesic that approaches
  the zero of $A(r,\tau)$ as $\tau \rightarrow \infty$
  defines the event horizon.
  To locate the position of this
  geodesic at early times, and hence determine the horizon area, 
  one must know the entire future of the geometry.
  }
Because of its acausal nature, the area of the event
horizon cannot correspond to the entropy of the system
in a non-equilibrium setting.
In contrast, it does appear sensible to regard the apparent
horizon area as a measure of thermodynamic entropy in a non-equilibrium
setting.

To facilitate a quantitative comparison between our numerical solutions
to Einstein's equations and the late time gradient expansion of
Ref.~\cite{Janik:2005zt},
Fig.~\ref{areagrad} shows a close-up view of the areas (per unit rapidity)
of the event and apparent horizons,
together with the corresponding late-time asymptotic expansions,
computed through second order in gradients.
These asymptotic results are
\cite{Kinoshita:2008dq, Nakamura:2006ih,Figueras:2009iu,HellerTBA1} 
\begin{subequations}
\label{asmarea}
\begin{align}
A_{\rm EH} & = \pi^3 \Lambda^2 \left [1 - \frac{1}{2\pi (\Lambda \tau)^{2/3}}  
+\frac{ 6 + \pi + 6 \ln 2}{24 \pi^2 (\Lambda \tau)^{4/3}} \right ],
\\
A_{\rm AH} & = \pi^3 \Lambda^2 \left [1 - \frac{1}{2\pi (\Lambda \tau)^{2/3}}  
+\frac{ 2 + \pi + \ln 2}{24 \pi^2 (\Lambda \tau)^{4/3}}\right ],
\end{align}
\end{subequations}
for the event and apparent horizon areas, respectively,
up to $\mathcal O \left ( (\Lambda \tau)^{-2} \right )$ corrections. 
{}From the figure one sees that the asymptotic expansions,
shown in the figure as the dashed black lines,
agree very well with the complete numerical results.
In fact, at time $\tau_f$
when the boundary geometry becomes flat, 
the asymptotic forms agree with the full numerical results for both
horizon areas to within $0.11\%$.

For $c = 1$, our numerically measured value of $\Lambda$ is $0.8$.
Consequently the first order corrections appearing in
Eqs.~(\ref{asmarea}) generate $10\%$ corrections at time $\tau_f$,
while the second order terms yield $0.20 \%$ and $0.56\%$ corrections
to the event and apparent horizon areas, respectively.

This comparison shows that the geometry in the bulk
(as probed by the horizon areas)
is already very well approximated by the 
gradient expansion of Ref.~\cite{Janik:2005zt} at time $\tau_f$.
However, it must be stressed that this very early agreement
with hydrodynamics is specific to the horizon areas, and is not
so true of other observables which are sensitive to the anisotropy
in the geometry, 
such as the SYM stress tensor,
which we discuss next.

\begin{figure*}[ht]
\includegraphics[scale=0.47]{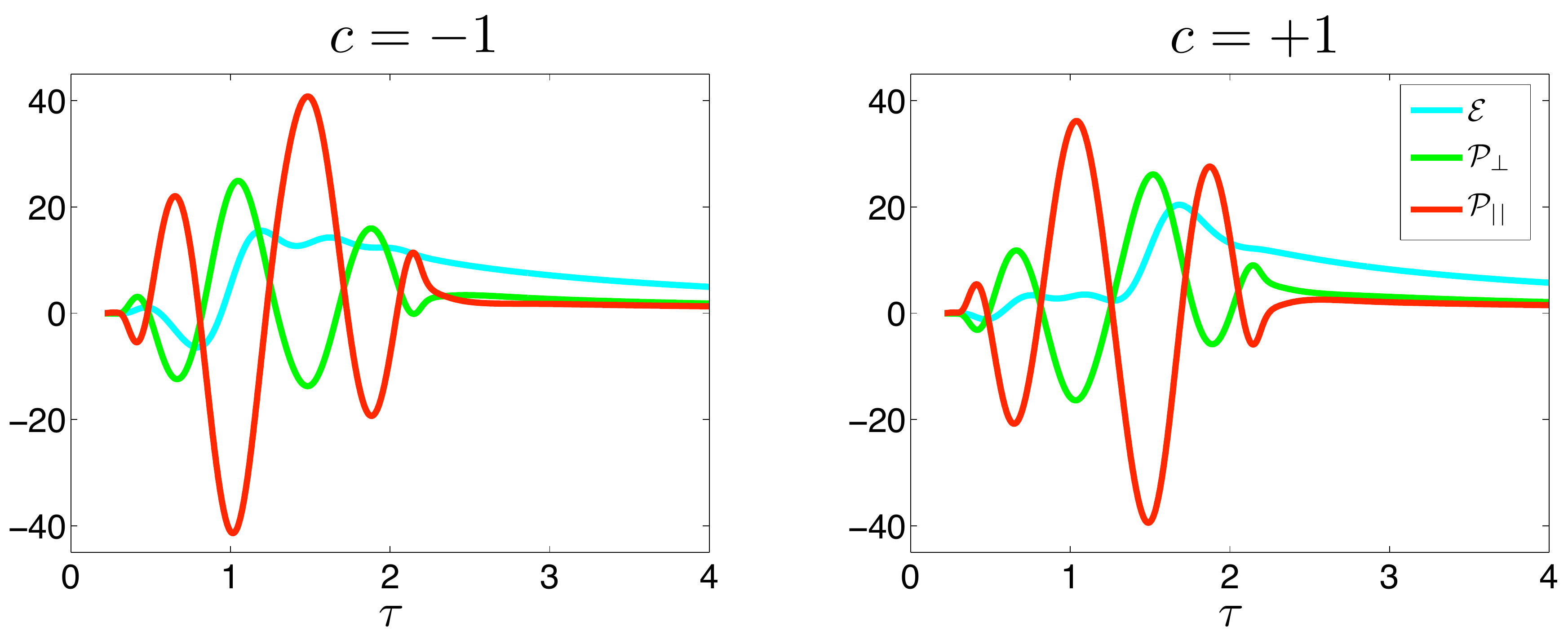}
\caption
    {\label{pressures}
   Energy density, 
   longitudinal
   and transverse pressure,
   all divided by $\Nc^2/2\pi^2$,
   as a function of time for $c = -1$ (left) and 
   $c = + 1$ (right).
   The energy density and pressures start off at zero at 
   time $\tau_i = 1/4$ when the system is in the vacuum state.
   During the interval of time $\tau \in (\tau_i,\tau_f) =  (0.25,2.25)$,
   the $4d$ geometry is changing and doing work on the field theory state.
   After time $\tau_f$ the deformation in the geometry turns off and the field
   theory state subsequently relaxes onto a hydrodynamic description.
   The smooth tails in both plots occur during this regime.
   At late times, from top to bottom, the three curves (in both plots)
   correspond to the
   energy density $\mathcal E$, transverse pressure $\mathcal P_\perp$,
   and longitudinal pressure $\mathcal P_\|$.
   }
\end{figure*}

We now turn to a discussion of our results for boundary field 
theory observables.
Fig.~\ref{pressures}
shows plots of the energy density 
and transverse and longitudinal pressures 
produced by the changing boundary geometry (\ref{boundarygeometry}),
when $c = \pm 1$.
These quantities begin at zero before time $\tau_i$,
when the system is in the vacuum state,
and deviate from zero once the $4d$ geometry starts to vary.
During the interval of time where the $4d$ geometry is changing,
the energy density generally grows and
the pressures rapidly oscillate: work is being done
on the field theory state.  After time $\tau_f$
the boundary geometry becomes flat and no longer does any work on the system.
As time progresses, non-hydrodynamic degrees of freedom relax and at
late times the evolution of the system is governed by hydrodynamics.
The late time hydrodynamic behavior manifests itself as the smooth
tails appearing in Fig.~\ref{pressures}.

The two sets of plots in Fig.~\ref{pressures}, contrasting $c = + 1$ and $-1$,
are qualitatively similar,
with the main difference being the phase of the oscillations
in the pressures.
For example, for $c = -1$
the transverse pressure is negative at $\tau_f$
whereas for $c = +1$
the transverse pressure is positive and larger than the longitudinal pressure,
which is nearly zero at $\tau_f$.  
As local equilibrium requires that the transverse and longitudinal pressure
be nearly equal \cite{Arnold:2004ti}, 
one sees that in either case the system is far from equilibrium at $\tau_f$.  
Furthermore, from the figure one sees that for either sign of $c$,
the transverse pressure approaches the longitudinal pressure from above.
As we next discuss,
this is always the case in the hydrodynamic limit of boost invariant flow.

{}From the gravitational asymptotic expansion of Ref.~\cite{Janik:2005zt},
one can compute the SYM stress tensor via Eq.~(\ref{stress}).
The results read \cite{Janik:2005zt}
\begin{subequations}
\label{hydrostress}
\begin{align}
\mathcal E &=  \frac{3 \pi^4 \Lambda^4}{4 (\Lambda \tau)^{4/3}}
\left [ 1 - \frac{2 C_1}{(\Lambda \tau)^{2/3}} + \frac{C_2}{(\Lambda \tau)^{4/3}}  \right],
\\
\mathcal P_ {\perp}&=  \frac{\pi^4 \Lambda^4}{4(\Lambda \tau)^{4/3}}
\left [ 1  - \frac{ C_2}{3 (\Lambda \tau)^{4/3}} \right],
\\
\mathcal P_ {||} &=   \frac{\pi^4 \Lambda^4}{4(\Lambda \tau)^{4/3}}
\left [ 1 - \frac{2 C_1}{(\Lambda \tau)^{2/3}} + \frac{5 C_2}{3 (\Lambda \tau)^{4/3}} \right],
\end{align} 
\end{subequations}
up to $\mathcal O((\Lambda \tau)^{-2})$ corrections.
The constant $C_1$ is related to the viscosity to entropy density ratio
of the plasma, while the constant $C_2$ is related to second-order
hydrodynamic relaxation times.
For strongly coupled SYM
\cite{Kinoshita:2008dq},
\begin{align}
C_1 = \frac{1}{3 \pi},  \ \ \ \ \ 
C_2 = \frac{2 + \ln 2}{18 \pi^2}.
\end{align}
The form (\ref{hydrostress}) for the stress-energy can also be obtained
from hydrodynamic considerations alone, together with knowledge of 
first and second order transport coefficients,
and the assumption of boost invariance \cite{Bjorken:1982qr,Baier:2007ix}.

\begin{figure*}[t!]
\includegraphics[scale=0.42]{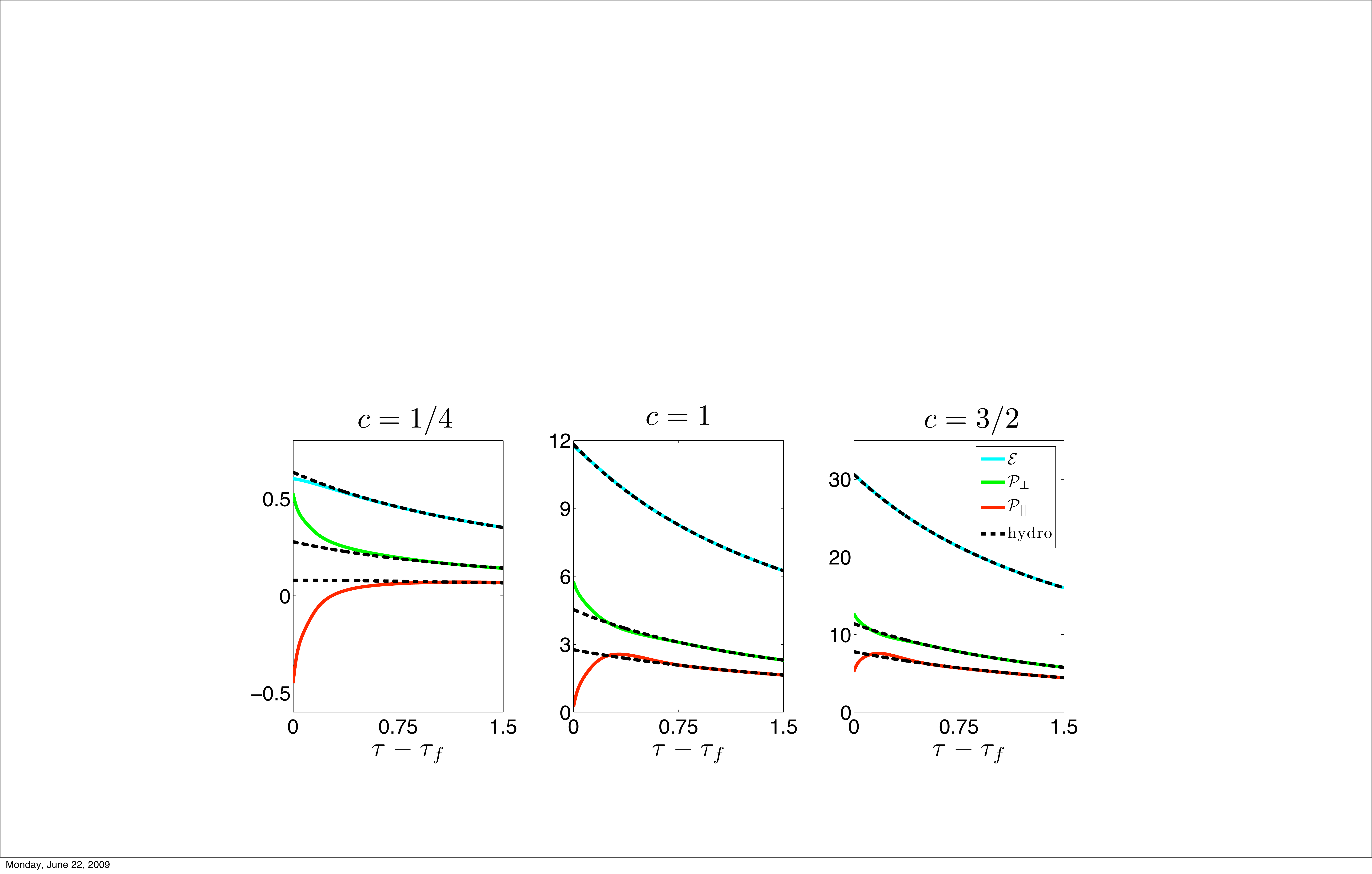}
\caption
    {\label{stresswithhydo}
    Energy density, 
    longitudinal
    and transverse pressure,
    all divided by $\Nc^2/2\pi^2$,
    as a function of time for $c = 1/4$ (left),
    $c = 1$ (middle) and $c = 3/2$ (right).    
    From top to bottom, the curves are energy density (blue),
    transverse pressure (green), and longitudinal pressure (red).
    The dashed black lines in each plot show the second order
    viscous hydrodynamic approximation (\ref{hydrostress}) to the different
    stress tensor components.
    Note the significantly different ordinate ranges in the three plots;
    the size of the difference between the transverse and longitudinal pressure
    grows with increasing $c$.
    }
\end{figure*}

It is evident from the leading terms of the result (\ref{hydrostress})
that at late times the 
stress-energy tensor approaches the ideal hydrodynamic form
\begin{equation}
\label{ideal}
T^{\mu}_{\ \nu} = \frac{\pi^2 N_c^2 T(\tau)^4}{8} \> {\rm diag} (-3,1,1,1),
\end{equation}
with a time-dependent temperature
\begin{equation}
\label{idealtemp}
T(\tau) =  \Lambda/(\Lambda \tau)^{1/3},
\end{equation}
which matches the Hawking temperature (\ref{hawking})
of the black brane in the gravitational description.
The ideal stress tensor (\ref{ideal}) is completely isotropic.
Subleading terms in the result (\ref{hydrostress}) show
that the transverse pressure differs from the longitudinal
pressure when viscous effects are taken into account.  In particular, as 
mentioned above,
first order viscous corrections make the transverse pressure larger
than the longitudinal pressure.

To facilitate a quantitative comparison
between our numerical results for the stress tensor and the
late-time hydrodynamic expansions, 
Fig.~\ref{stresswithhydo} shows the energy density and pressures
for $c = 1/4, \ 1$ and $3/2$, with the corresponding hydrodynamic 
forms (\ref{hydrostress}) plotted on top of the numerical data.
The plots start at time $\tau = \tau_f$.  In all three plots,
one clearly sees the stress-energy components approach their
hydrodynamic approximations.
Moreover, in all plots one sees a  substantial anisotropy even
at late times where a hydrodynamic treatment is applicable.
In other words, the effect of viscosity is
very evident in these results.

\begin{table*}[t!]
\begin{tabular}{@{\extracolsep{5pt}}c|cccccccccc}
$c$ & $-2$ & $-3/2$ & $-1$ &$-1/2$ & $-1/4$ & 1/4 & 1/2 & 1 & 3/2 & 2 
\\
\hline
\hline
$\tau_{*}$ & 2.2 & 2.3 & 2.4 & 2.7 & 3.1 & 3.1 & 2.7 & 2.4 & 2.3 & 2.2
\\[2pt]
$T_*$ & 0.93 & 0.77 & 0.60 & 0.40 & 0.27 & 0.27 & 0.41 & 0.62 & 0.80 & 0.97 
\\[2pt]
$\Lambda \tau_{*} $ & 3.1 & 2.5 & 1.9 & 1.2 & 0.87 & 0.89 & 1.3 & 1.9 & 2.6 & 3.3 
\\[2pt]
$( \tau_{*} {-} \tau_i) \, T_*$ & 2.0 & 1.7 & 1.4 & 1.1 & 0.84 & 0.85 & 1.1 & 1.5 & 1.8 & 2.1
\\[2pt]
$( \tau_{*} {-} \tau_f) \, T_*$ & 0.00 & 0.05 & 0.11 & 0.19 & 0.24 & 0.24 & 0.20 & 0.11 & 0.04 & 0.00 
\\[2pt]
$\frac{ \mathcal P_\perp(\tau_f) - P_{||}(\tau_f)}{\mathcal E(\tau_f)}$
 & 0.06 & $-0.03$ & $-0.22$ & $-0.56$ & $-1.1$ & 1.6 & 0.91 & 0.47 & 0.24 & 0.13 
\end{tabular}
\caption
    {%
    Quantities characterizing the relaxation to equilibrium,
    for various values of the boundary perturbation amplitude $c$.
    The relaxation time $\tau_{*}$
    (in units of $\Delta$)
    is the time at which the transverse
    and longitudinal pressures deviate from their hydrodynamic values
    (\ref{hydrostress}) by less than 10\%.
    $T_*$ is the temperature at time $\tau_*$,
    and $\Lambda$ is the scale appearing in
    the hydrodynamic expansion (\ref{hydrostress})
    (both measured in units of $\Delta^{-1}$).
    The quantity $( \tau_{*} {-} \tau_i) \, T_*$
    measures the total time in units of $T_*$ required to produce the
    plasma and relax to near local-equilibrium.
    The quantity $( \tau_{*} {-} \tau_f) \, T_*$
    measures the time in units of $T_*$ required
    for the plasma to relax after the deformation in the geometry ceases.
    The quantity
    $
	\left[ \mathcal P_\perp(\tau_f) - P_{||}(\tau_f) \right]
	/{\mathcal E(\tau_f)}
    $
    is the pressure anisotropy, relative to the energy density,
    at time $\tau_f$.
    }
\label{T1}
\end{table*}

Looking at Fig.~\ref{stresswithhydo}, for time $\tau = \tau_f$ and $c = 1/4$,
one sees that the transverse and longitudinal pressures are almost 
equal and opposite in magnitude at this time.  So the system
is initially very far-from-equilibrium.   However, for $c = 3/2$
the pressures are both positive,
and system is much closer to equilibrium at $\tau_f$.
At first sight this might seem peculiar: 
how can it be that for larger values of $c$,
where the size of the perturbation in the $4d$ geometry is huge,
the system takes less time to reach local equilibrium!
Qualitatively, this apparent puzzle is easy to understand.
For large $c$, the changing geometry 
does more work on the system and consequently the system
reaches a higher effective temperature.
Because SYM is a conformal theory, relaxation times for non-hydrodynamic 
degrees of freedom must scale inversely with the temperature,
and hence must vanish as the local energy density diverges. 
Therefore, in the $c \rightarrow \infty$ limit the system will
always be very close to local equilibrium --- even while the $4d$
geometry is changing --- and the anisotropy in the pressures
will vanish immediately at $\tau_f$.  
As a consequence, one learns little about the physics of the 
relaxation of non-hydrodynamic degrees of freedom in the 
$c \rightarrow \infty$ limit.  

Table~\ref{T1} shows how various quantities characterizing
the relaxation of the plasma depend on the boundary perturbation
amplitude $c$, within the range $[-2,2]$.
Included in the table is
the time $\tau_*$, beyond which the stress tensor
agrees with the hydrodynamic approximation (\ref{hydrostress})
to within $10\%$.  
Also shown is the temperatures $T_*$ at time $\tau_*$,
the scale $\Lambda$ measured in units of $\tau_*$,
and the time intervals $\tau_* {-} \tau_i$
and $\tau_* {-} \tau_f$ measured in units of $T_*$.  

{}From the table, one sees that 
as the magnitude of $c$ increases, so does the temperature $T_*$.
Moreover, as the magnitude of $c$ increases, one sees that
the time scale $\tau_*$ approaches $\tau_f = 2.25$.
In particular, for $|c| = 2$ the stress tensor is already
within $10\%$ of its hydrodynamic limit at $\tau_f$.
As discussed above, both of these features are to be expected.
Increasing $|c|$ means that the changing geometry does more
work on the system, producing a larger energy density,
and consequently the relaxation times of non-hydrodynamic
degrees of freedom decrease.
In all cases presented in Table~\ref{T1},
the relevant dynamics --- from the production of the plasma to
its relaxation to near local equilibrium (where hydrodynamics applies) ---
occur over a time $\tau_* - \tau_i \lesssim 2/T_*$.

{}From Table~\ref{T1}, one also sees that for $|c| \lesssim 1/2$
the time scale $\tau_*$ at which a hydrodynamic treatment becomes
accurate occurs when $\Lambda \tau_*  \approx 1$.
For larger values of $|c|$, $\Lambda \tau_*$ is bigger.
Examining the size of the coefficients in
Eq.~(\ref{hydrostress}) shows that
the hydrodynamic expansion is quite well-behaved
for $\Lambda \tau \gtrsim 1$.
{}From the time-temperature relation (\ref{idealtemp}),
one can convert $\Lambda \tau_* \gtrsim 1$ to the estimate
$\tau_* T_* \gtrsim 1$ which, from Table~\ref{T1}\,, is indeed the case.

Examining the size of the cofficients in
the series (\ref{hydrostress})
shows that the second-order $(\Lambda \tau)^{-4/3}$ terms are quite
small compared to the leading $(\Lambda\tau)^{-2/3}$ viscous terms
when $\Lambda\tau \ge 1$;
they only become comparable when $\Lambda\tau$ is below 0.1.
Hence, the fact that hydrodynamics is not accurate until
$\Lambda\tau $ is larger than 1--2 (depending on the value of $c$)
indicates that the physics which determines the onset of hydrodynamic behavior
is not responsible for higher order terms in the hydrodynamic
expansion becoming comparable to lower order terms.
Rather, the change in behavior from non-hydrodynamic far-from-equilibrium
behavior to near-local-equilibrium hydrodynamic response must be
reflecting the relative importance of exponentially relaxing
non-hydrodynamic degrees of freedom in comparison to the
slowly relaxing hydrodynamic modes.
This means that one cannot accurately identify the domain of utility of
the hydrodynamic description by asking when the late-time gradient
expansion breaks down.  A similar conclusion was also reached
in Ref.~\cite{Amado:2008ji} by analyzing small perturbations
on top of an infinite static plasma.

It is instructive to discuss the qualitative origin of the
relaxation time $\tau_*$ from the perspective of the $5d$ gravitational problem.
First consider the limit $|c| \rightarrow \infty$.
In this limit, large amounts of gravitational radiation
are produced by the changing boundary geometry 
and the amount of energy which falls deep into the bulk diverges.  
As a consequence, the horizon radius must approach the 
boundary as $|c| \to \infty$.
The infall time for radiation to travel from the boundary
to the horizon is roughly equal to the inverse horizon radius
(in our coordinate system),
so in the $|c| \rightarrow \infty$ limit,
the geometry outside the horizon effectively
responds instantaneously to the changing boundary geometry.
Therefore, as $|c| \rightarrow \infty$,
the system needs no to time to return to local equilibrium
after the geometry stops changing at time $\tau_f$.

Now consider the $|c| \rightarrow 0$ limit.
For small $|c|$, distinct dynamics occurs on the time scales
$\Delta$ and $\Delta/\sqrt{|c|}$
[with $\Delta \equiv \frac{1}{2} (\tau_f - \tau_i)$].
First of all, irrespective of how small $c$ is,
the positions of the apparent and event horizons are rapidly
varying only over the time scale $\Delta$.
This is because it is during the time interval
$\tau_i \le \tau \le \tau_f$ that gravitational
radiation is being produced and absorbed by the horizon,
creating most growth in horizon area.
The parametric size of the horizon radii at $\tau_f$
is $\sim \sqrt{|c|}/\Delta$.
Qualitatively, this makes sense since
little radiation is produced and very little radiation
falls into the bulk of the geometry when $|c|$ is small.
Hence, the black hole size will vanish as $|c|\to 0$.
An easy way to understand the $\sqrt{|c|}$ scaling is to note that the
total energy added to the field theory state cannot depend on the sign
of $c$, and therefore must be quadratic in $c$ in the small $|c|$ limit.
The final state energy density in the field theory will scale as
$r_h^4$, so the horizon radii must be proportional to $\sqrt{|c|}$.
Consequently, after time $\tau_f$ it takes a time
$\sim \Delta/\sqrt{|c|}$ for any remaining short wavelength
perturbations to fall into the horizon.
It is during this interval of time that the geometry
undergoes its relaxation onto the slowly evolving hydrodynamic form.
We therefore see that $\tau_* T_*$ should have a non-zero
$\mathcal O(1)$ limit when $|c| \rightarrow 0$.%
\footnote
    {%
    A brief comment on the relation between our work and
    the recent paper of Bhattacharyya and Minwalla \cite {Bhattacharyya:2009uu}
    may be in order.
    These authors examined black hole formation and thermalization
    in response to an arbitrarily weak boundary perturbation coupling
    to the dilaton.
    A noteworthy finding in this work was ``instant thermalization''
    (as probed by measurements of local operators)
    after the boundary perturbation turned off.
    However, this is the case for asymptotically AdS$_4$ spacetime and,
    as clearly stated in Ref.~\cite {Bhattacharyya:2009uu},
    is not expected to hold in asymptotically AdS$_5$ spacetime
    or for non-infinitesimal boundary perturbations.
    }

Last, we discuss the relevance of our work to more complicated numerical
relativity problems in gauge/gravity duality.
As discussed in the Introduction,
an interesting future direction is the study of
collisions of gravitational shock waves in AdS$_5$,
as this is dual to the collision of sheets of matter in SYM and
mimics the collision of large, highly boosted nuclei in heavy ion collisions.
In the simplest setting, one can study shock waves which are translationally
invariant in two transverse directions \cite{Grumiller:2008va}.
The corresponding gravitational problem is therefore $2+1$ dimensional.
While we have studied a simpler $1+1$ dimensional gravitational problem
in this paper, there are several lessons which may provide
insight relevant for more difficult problems.
First, we found it necessary to solve Einstein's equations
analytically near the boundary with a power series expansion in the
radial coordinate.
This was required as the presence of the negative cosmological constant
makes the near-boundary geometry singular.
More specifically, careful asymptotic near-boundary analysis was
required to determine the appropriate subtraction terms needed
to make the integrals (\ref{ThetaPhi}) finite and produce a
numerical scheme which remains accurate near the boundary.
The same issue will arise in gravitational problems
with less symmetry.

Another important lesson concerns the choice of coordinates.
Because of the presence of the negative cosmological constant,
all matter and radiation tends to fall inward into the bulk.
This universality of gravitational infall motivates the use of
coordinates specifically adapted for infalling motion.
The generalized infalling Eddington-Finkelstein
coordinates we used, which assign a constant ``time'' coordinate
to all events on infalling null radial geodesics,
are especially appropriate for numerics.
Had we used a time coordinate which defined a spacelike slicing
of the geometry, then we would have wasted computational time
solving for the geometry deep in the bulk before any signals from
the boundary had arrived.
With a null time coordinate,
a signal propagating in from the boundary at $r = \infty$
arrives ``\textit{instantaneously}'' at $r = 0$.
Moreover, the generalized infalling Eddington-Finkelstein coordinates
yield a metric (\ref{metric}) which is non-singular on the horizon.
Coordinates which do not yield a metric regular at the horizon,
such as Fefferman-Graham coordinates,
are not well suited to numerical initial value problems.

\section{Conclusions}

Using gauge/gravity duality, we have studied the production and
relaxation of a boost invariant plasma in strongly coupled
$\Nfour$ supersymmetric Yang-Mills theory.
The production mechanism is a time-dependent deformation of
the four dimensional geometry in which the field theory lives.
The deformation, which was confined to a compact interval
of proper time, does work on the system and thus excites
the initial state, which we took to the $\Nfour$ SYM vacuum.
Within the context of gauge/gravity duality, this problem
maps into the problem of black hole formation in five dimensions.
By solving the corresponding gravitational problem numerically, and
using the gauge/gravity dictionary, we were able to compute the
field theory stress tensor at all times, from the first excitation
of the initial vacuum state to the late-time onset of hydrodynamics.
We found that the entire process of plasma creation --- from the initial
vacuum state to the relaxation onto a hydrodynamic description ---
can occur in times as short as one to two times $1/T_*$, where $T_*$
is the local temperature at the onset of the hydrodynamic regime.
We also demonstrated that the time at which a hydrodynamic treatment
first becomes valid does not coincide with the point where the
hydrodynamic gradient expansion breaks down.
This reflects the fact that in a far-from-equilibrium state
there are non-hydrodynamic degrees of freedom.
These modes relax exponentially, and their relative importance
determines the onset of the hydrodynamic regime.

This work, together with our earlier paper \cite{Chesler:2008hg},
provide novel additions to the very sparse set of examples of genuinely
far-from-equilibrium processes in quantum field theory which can be
studied with complete theoretical control.  Using techniques
similar to those presented in this paper,
it should be possible to study more
demanding problems which have less symmetry.

\begin{acknowledgments}
This work was supported in part by the U.S. Department
of Energy under Grant No.~DE-FG02-\-96ER\-40956.
We are grateful to
Michal Heller, Andreas Karch, and Paul Romatschke
for useful discussions.
\end{acknowledgments}


\bibliographystyle{utphys}
\bibliography{refs}%

\providecommand{\href}[2]{#2}\begingroup\raggedright\begin{thebibliography}{10}

\bibitem{Arnold:2002zm}
P.~Arnold, G.~D. Moore, and L.~G. Yaffe, ``{Effective kinetic theory for high
  temperature gauge theories},'' {\em JHEP} {\bf 01} (2003)  030,
\href{http://arxiv.org/abs/hep-ph/0209353}{{\tt arXiv:hep-ph/0209353}}.

\bibitem{Arnold:2003zc}
P.~Arnold, G.~D. Moore, and L.~G. Yaffe, ``{Transport coefficients in high
  temperature gauge theories. II: Beyond leading log},'' {\em JHEP} {\bf 05}
  (2003)  051,
\href{http://arxiv.org/abs/hep-ph/0302165}{{\tt arXiv:hep-ph/0302165}}.

\bibitem{Arnold:2004ti}
P.~Arnold, J.~Lenaghan, G.~D. Moore, and L.~G. Yaffe, ``{Apparent
  thermalization due to plasma instabilities in quark gluon plasma},''
  \href{http://dx.doi.org/10.1103/PhysRevLett.94.072302}{{\em Phys. Rev. Lett.}
  {\bf 94} (2005)  072302},
\href{http://arxiv.org/abs/nucl-th/0409068}{{\tt arXiv:nucl-th/0409068}}.

\bibitem{Arnold:2000dr}
P.~Arnold, G.~D. Moore, and L.~G. Yaffe, ``{Transport coefficients in high
  temperature gauge theories: (I) Leading-log results},'' {\em JHEP} {\bf 11}
  (2000)  001,
\href{http://arxiv.org/abs/hep-ph/0010177}{{\tt arXiv:hep-ph/0010177}}.

\bibitem{Jeon:1995zm}
S.~Jeon and L.~G. Yaffe, ``{From quantum field theory to hydrodynamics:
  Transport coefficients and effective kinetic theory},''
  \href{http://dx.doi.org/10.1103/PhysRevD.53.5799}{{\em Phys. Rev.} {\bf D53}
  (1996)  5799--5809},
\href{http://arxiv.org/abs/hep-ph/9512263}{{\tt arXiv:hep-ph/9512263}}.

\bibitem{Kovtun:2004de}
P.~Kovtun, D.~T. Son, and A.~O. Starinets, ``{Viscosity in strongly interacting
  quantum field theories from black hole physics},''
  \href{http://dx.doi.org/10.1103/PhysRevLett.94.111601}{{\em Phys. Rev. Lett.}
  {\bf 94} (2005)  111601},
\href{http://arxiv.org/abs/hep-th/0405231}{{\tt arXiv:hep-th/0405231}}.

\bibitem{Bhattacharyya:2008jc}
S.~Bhattacharyya, V.~E. Hubeny, S.~Minwalla, and M.~Rangamani, ``{Nonlinear
  fluid dynamics from gravity},''
  \href{http://dx.doi.org/10.1088/1126-6708/2008/02/045}{{\em JHEP} {\bf 02}
  (2008)  045},
\href{http://arxiv.org/abs/0712.2456}{{\tt arXiv:0712.2456 [hep-th]}}.

\bibitem{Chesler:2007sv}
P.~M. Chesler and L.~G. Yaffe, ``{The stress-energy tensor of a quark moving
  through a strongly-coupled $\mathcal N=4$ supersymmetric Yang-Mills plasma:
  comparing hydrodynamics and AdS/CFT},''
  \href{http://dx.doi.org/10.1103/PhysRevD.78.045013}{{\em Phys. Rev.} {\bf
  D78} (2008)  045013},
\href{http://arxiv.org/abs/0712.0050}{{\tt arXiv:0712.0050 [hep-th]}}.

\bibitem{Kovtun:2005ev}
P.~K. Kovtun and A.~O. Starinets, ``{Quasinormal modes and holography},''
  \href{http://dx.doi.org/10.1103/PhysRevD.72.086009}{{\em Phys. Rev.} {\bf
  D72} (2005)  086009},
\href{http://arxiv.org/abs/hep-th/0506184}{{\tt arXiv:hep-th/0506184}}.

\bibitem{Maldacena:1997re}
J.~M. Maldacena, ``The large {$N$} limit of superconformal field theories and
  supergravity,'' {\em Adv. Theor. Math. Phys.} {\bf 2} (1998)  231--252,
\href{http://arxiv.org/abs/hep-th/9711200}{{\tt arXiv:hep-th/9711200}}.

\bibitem{Witten:1998qj}
E.~Witten, ``{Anti-de Sitter} space and holography,'' {\em Adv. Theor. Math.
  Phys.} {\bf 2} (1998)  253--291,
\href{http://arxiv.org/abs/hep-th/9802150}{{\tt arXiv:hep-th/9802150}}.

\bibitem{GKP}
S.~S. Gubser, I.~R. Klebanov, and A.~M. Polyakov, ``Gauge theory correlators
  from non-critical string theory,'' {\em Phys. Lett.} {\bf B428} (1998)
  105--114,
\href{http://arxiv.org/abs/hep-th/9802109}{{\tt hep-th/9802109}}.

\bibitem{Shuryak}
E.~Shuryak, ``Why does the quark gluon plasma at {RHIC} behave as a nearly
  ideal fluid?,'' {\em Prog. Part. Nucl. Phys.} {\bf 53} (2004)  273--303,
\href{http://arxiv.org/abs/hep-ph/0312227}{{\tt hep-ph/0312227}}.

\bibitem{Shuryak:2004cy}
E.~V. Shuryak, ``{What RHIC experiments and theory tell us about properties of
  quark-gluon plasma?},''
  \href{http://dx.doi.org/10.1016/j.nuclphysa.2004.10.022}{{\em Nucl. Phys.}
  {\bf A750} (2005)  64--83},
\href{http://arxiv.org/abs/hep-ph/0405066}{{\tt arXiv:hep-ph/0405066}}.

\bibitem{Heinz:2004pj}
U.~W. Heinz, ``Thermalization at {RHIC},''
  \href{http://dx.doi.org/10.1063/1.1843595}{{\em AIP Conf. Proc.} {\bf 739}
  (2005)  163--180},
\href{http://arxiv.org/abs/nucl-th/0407067}{{\tt arXiv:nucl-th/0407067}}.

\bibitem{Grumiller:2008va}
D.~Grumiller and P.~Romatschke, ``On the collision of two shock waves in
  {AdS$_5$},'' \href{http://dx.doi.org/10.1088/1126-6708/2008/08/027}{{\em
  JHEP} {\bf 08} (2008)  027},
\href{http://arxiv.org/abs/0803.3226}{{\tt arXiv:0803.3226 [hep-th]}}.

\bibitem{Gubser:2008pc}
S.~S. Gubser, S.~S. Pufu, and A.~Yarom, ``{Entropy production in collisions of
  gravitational shock waves and of heavy ions},''
  \href{http://dx.doi.org/10.1103/PhysRevD.78.066014}{{\em Phys. Rev.} {\bf
  D78} (2008)  066014},
\href{http://arxiv.org/abs/0805.1551}{{\tt arXiv:0805.1551 [hep-th]}}.

\bibitem{AlvarezGaume:2008fx}
L.~Alvarez-Gaume, C.~Gomez, A.~S. Vera, A.~Tavanfar, and M.~A. Vazquez-Mozo,
  ``{Critical formation of trapped surfaces in the collision of gravitational
  shock waves},''
\href{http://arxiv.org/abs/0811.3969}{{\tt arXiv:0811.3969 [hep-th]}}.

\bibitem{Lin:2009pn}
S.~Lin and E.~Shuryak, ``{Grazing collisions of gravitational shock waves and
  entropy production in heavy ion collision},''
  \href{http://arxiv.org/abs/0902.1508}{{\tt arXiv:0902.1508 [hep-th]}}.

\bibitem{Gubser:2009sx}
S.~S. Gubser, S.~S. Pufu, and A.~Yarom, ``{Off-center collisions in $AdS_5$
  with applications to multiplicity estimates in heavy-ion collisions},''
\href{http://arxiv.org/abs/0902.4062}{{\tt arXiv:0902.4062 [hep-th]}}.

\bibitem{Chesler:2008hg}
P.~M. Chesler and L.~G. Yaffe, ``{Horizon formation and far-from-equilibrium
  isotropization in supersymmetric Yang-Mills plasma},''
\href{http://arxiv.org/abs/0812.2053}{{\tt arXiv:0812.2053 [hep-th]}}.

\bibitem{Bhattacharyya:2009uu}
S.~Bhattacharyya and S.~Minwalla, ``{Weak field black hole formation in
  asymptotically AdS spacetimes},''
\href{http://arxiv.org/abs/0904.0464}{{\tt arXiv:0904.0464 [hep-th]}}.

\bibitem{Birrell:1982ix}
N.~D. Birrell and P.~C.~W. Davies, ``{Quantum Fields in Curved Space},''.
  Cambridge Univ. Pr., UK.

\bibitem{Janik:2005zt}
R.~A. Janik and R.~B. Peschanski, ``{Asymptotic perfect fluid dynamics as a
  consequence of AdS/CFT},''
  \href{http://dx.doi.org/10.1103/PhysRevD.73.045013}{{\em Phys. Rev.} {\bf
  D73} (2006)  045013},
\href{http://arxiv.org/abs/hep-th/0512162}{{\tt arXiv:hep-th/0512162}}.

\bibitem{Kinoshita:2008dq}
S.~Kinoshita, S.~Mukohyama, S.~Nakamura, and K.-y. Oda, ``{A holographic dual
  of Bjorken flow},'' \href{http://dx.doi.org/10.1143/PTP.121.121}{{\em Prog.
  Theor. Phys.} {\bf 121} (2009)  121--164},
\href{http://arxiv.org/abs/0807.3797}{{\tt arXiv:0807.3797 [hep-th]}}.

\bibitem{Heller:2009zz}
M.~P. Heller, P.~Surowka, R.~Loganayagam, M.~Spalinski, and S.~E. Vazquez,
  ``{Consistent holographic description of boost-invariant plasma},''
  \href{http://dx.doi.org/10.1103/PhysRevLett.102.041601}{{\em Phys. Rev.
  Lett.} {\bf 102} (2009)  041601},
\href{http://arxiv.org/abs/0805.3774}{{\tt arXiv:0805.3774 [hep-th]}}.

\bibitem{janikandheler}
G.~Beuf, M.~P. Heller, R.~A. Janik, and R.~Peschanski, ``{Boost-invariant early
  time dynamics from AdS/CFT},'' \href{http://arxiv.org/abs/0906.4423}{{\tt
  arXiv:0906.4423 [hep-th]}}.

\bibitem{deHaro:2000xn}
S.~de~Haro, S.~N. Solodukhin, and K.~Skenderis, ``Holographic reconstruction of
  spacetime and renormalization in the {AdS/CFT} correspondence,''
  \href{http://dx.doi.org/10.1007/s002200100381}{{\em Commun. Math. Phys.} {\bf
  217} (2001)  595--622},
\href{http://arxiv.org/abs/hep-th/0002230}{{\tt arXiv:hep-th/0002230}}.

\bibitem{Anninos:1994dj}
P.~Anninos, G.~Daues, J.~Masso, E.~Seidel, and W.-M. Suen, ``Horizon boundary
  condition for black hole space-times,''
  \href{http://dx.doi.org/10.1103/PhysRevD.51.5562}{{\em Phys. Rev.} {\bf D51}
  (1995)  5562--5578},
\href{http://arxiv.org/abs/gr-qc/9412069}{{\tt arXiv:gr-qc/9412069}}.

\bibitem{Wald:1984rg}
R.~M. Wald, ``General relativity,''. {Chicago, Usa: Univ. Pr. 491p}.

\bibitem{Nakamura:2006ih}
S.~Nakamura and S.-J. Sin, ``{A holographic dual of hydrodynamics},'' {\em
  JHEP} {\bf 09} (2006)  020,
\href{http://arxiv.org/abs/hep-th/0607123}{{\tt arXiv:hep-th/0607123}}.

\bibitem{Figueras:2009iu}
P.~Figueras, V.~E. Hubeny, M.~Rangamani, and S.~F. Ross, ``{Dynamical black
  holes and expanding plasmas},''
\href{http://arxiv.org/abs/0902.4696}{{\tt arXiv:0902.4696 [hep-th]}}.

\bibitem{HellerTBA1}
I.~Booth, M.~P. Heller, and M.~Spalinski, ``{Gravity dual to boost-invariant
  flow as a slowly-evolving geometry},''. To appear shortly.

\bibitem{Bjorken:1982qr}
J.~D. Bjorken, ``{Highly relativistic nucleus-nucleus collisions: The central
  rapidity region},''
\href{http://dx.doi.org/10.1103/PhysRevD.27.140}{{\em Phys. Rev.} {\bf D27}
  (1983)  140--151}.

\bibitem{Baier:2007ix}
R.~Baier, P.~Romatschke, D.~T. Son, A.~O. Starinets, and M.~A. Stephanov,
  ``{Relativistic viscous hydrodynamics, conformal invariance, and
  holography},'' \href{http://dx.doi.org/10.1088/1126-6708/2008/04/100}{{\em
  JHEP} {\bf 04} (2008)  100},
\href{http://arxiv.org/abs/0712.2451}{{\tt arXiv:0712.2451 [hep-th]}}.

\bibitem{Amado:2008ji}
I.~Amado, C.~Hoyos-Badajoz, K.~Landsteiner, and S.~Montero, ``{Hydrodynamics
  and beyond in the strongly coupled N=4 plasma},''
  \href{http://dx.doi.org/10.1088/1126-6708/2008/07/133}{{\em JHEP} (2008)
  133},
\href{http://arxiv.org/abs/0805.2570}{{\tt arXiv:0805.2570 [hep-th]}}.

\end{thebibliography}\endgroup
\end{document}